\documentstyle[12pt,epsf]{article}

\begin{document}

\title{Low temperature kinetics of 2D exciton gas cooling in 
quantum well bilayer}
\author{S.V.Iordanski and A.Kashuba}

\maketitle

\vskip 0.5in

We study the kinetics of 2D Bose gas cooling provided Bose particles 
interact with 3D phonons. At low temperatures phonon emission is 
prohibited by the energy and the momentum conservation. We show that 
both particle-particle scattering and impurity scattering assist the
Bose gas cooling. The temporal relaxation of temperature follows the 
law $T\sim 1/\sqrt{t}$ above the Berezinski-Kosterlitz-Thouless phase 
transition point and $T\sim 1/t$ after a Bose-Einstein 2D 
quasi-condensate develops.

\vskip 0.3in

Exciton gas in GaAs bilayer quantum well represents a system where a 2D 
Bose-Einstein quasi-condensation (BEqC) is possible at low temperatures. 
Experimental efforts \cite{BF98,LTHS00} have been directed to produce 
such an exciton gas and to cool it down to a BEqC temperature. A short 
laser photo-illumination pulse excites electron and hole pairs. The 
so-called indirect exciton technique is used where a perpendicular 
electric field drags electrons and holes apart into two spatially 
separated layers. Then an electron and a hole bind themselves into an 
indirect exciton particle. This experimental setup suppresses the 
electron-hole recombination giving rise to a relatively long exciton 
life time. During the initial photoexcitation pulse newly born excitons 
are hot and form a non-equilibrium state. In short time after the pulse 
ends the exciton gas reaches an equilibrium at some effective exciton 
temperature which is much higher than the lattice temperature of the 
cold GaAs crystal. Frequent exciton-exciton collisions ensures the 
exciton temperature to be uniform across the bilayer. Exciton gas then 
start to cool down slowly due to an emission of phonons into a crystal 
away from the bilayer. This is the longest phase of the experiment 
limited only by a decay time of excitons due to the electron-hole 
recombination. In order to reach a BEqC point one needs both low 
temperature and a high density of excitons. Hence a fast cooling is 
essential.

Recent calculation of energy losses in 2D ideal exciton system has
predicted an extremely slow cooling at low temperatures with the 
temporal law: $T(t)\sim 1/\log(t)$, where $t$ is the time \cite{ILH98}. 
This fact is intimately related to the energy and momentum conservation 
which prohibits an emission of phonons by an exciton moving slower 
than the velocity of sound in GaAs crystal: $c$. Thus, the exciton 
gas cooling in effect stops when the exciton temperature falls below 
a characteristic blocking temperature: $T_b=mc^2/2$, where $m$ is the 
mass of exciton, even if the crystal temperature is zero.

This kinetic bottleneck problem becomes especially acute when the
exciton gas is subjected to a strong perpendicular magnetic field 
that quenches the motion of exciton to the lowest Landau level and, 
thus, is helping to bind electrons and holes into exciton pairs. In 
this case the effective mass of an exciton is determined by the 
Coulomb interaction and can be much larger than either the electron 
or hole mass \cite{LR97}. This results in a higher blocking 
temperature $T_b$ and makes it difficult to reach low temperatures 
in the end.

In this communication we supplement the analysis of exciton cooling 
of Ref.\cite{ILH98} by an addition of exciton-exciton collisions and 
scattering on impurities. Both events substantially assist the phonon 
emission. We specialize to the case of exactly zero lattice 
temperature which allows us to neglect exciton-phonon scattering. 
Actually we are dealing with a general problem of 2D Bose gas cooling
provided its particles interact with 3D phonons. The universal nature 
of 2D scattering at low energy of incoming particles makes these two 
assistance mechanisms to be robust to specific details of a 
particle-particle or impurity potential. The latter is only assumed 
to be short-ranged with the characteristic interaction radius $r_0$ 
being shorter than the DeBroigle wavelength. For example, an indirect 
exciton interacts with an impurity or another exciton via the 
electron-hole dipole moment: $e\vec{d}$, directed along the normal 
to the bilayer. Hence, the radius of such a dipole interaction equals
to the spacing between the electron and the hole layers: $r_0\sim d$.

The Hamiltonian of particle phonon interaction can be written 
generally as:
\begin{equation}\label{Ex-PhHam} H_{x-ph}=\int\int \psi^+(\vec{r})
\psi(\vec{r})\delta(z)\Gamma_i(\vec{\rho}-\vec{\rho'})u_i(\vec{\rho'})
\ d^3\vec{\rho'} d^2\vec{r}dz, \end{equation} where $\vec{\rho}=
(\vec{r},z)$, $\psi^+$ and $\psi$ are the particle creation and 
annihilation operators and $\vec{u}$ is a crystal deformation induced 
by an acoustic phonon. Wavelengths of relevant phonons are much larger 
than the width of the bilayer: $d$. The lattice deformation $\vec{u}$ 
can be expanded into the normal phonon modes as: 
\begin{equation}\label{PhModes}u_i(\vec{\rho})=\sum_{s, \vec{q}}\left(
{\hbar\over 2\rho\omega_s(\vec{q})}\right)^{1/2}(e^s_i b^+_s(-\vec{q})+
e^{*s}_i b_s(\vec{q})) e^{i\vec{q}\vec{\rho}},\end{equation} where 
$b^+_s$ and $b_s$ are the phonon creation and annihilation operators 
of polarization $s$, $\rho$ is the mass density of solid, $\omega_s
(\vec{q})=cq$ is the phonon frequency dispersion, which we assume to 
be isotropic and independent of phonon polarization $s$. 

The exciton phonon interaction in GaAs crystal can be separated into
piezoelectric and deformation potential parts. Lattice deformation in
a piezoelectric crystal induces a polarization density: $P_i=\beta_{ijk}
\partial_j u_k$ \cite{GL}, where $\beta_{ijk}$ is the piezoelectric 
tensor. This polarization interacts with the exciton dipole moment. In 
the limit $qd \ll 1$, the deformation potential for an exciton: 
$\Theta$, is a sum of the deformation potentials for an electron and a 
hole taken at the same spatial point. The latter represents a change of 
the semiconductor gap due to the local compression caused by a phonon 
deformation. Combining the piezoelectric and deformation parts and 
expanding the crystal lattice deformation in acoustic phonon modes we 
write the the exciton-phonon vertex in the Hamiltonian (\ref{Ex-PhHam}) 
as \cite{GL}:
\begin{equation}\label{GammaPiez}\Gamma_i(\vec{q})=\left(\Theta q_i+ed
\beta_{ijk}{4\pi q_zq_jq_k\over\vec{q}^2}\right). \end{equation} For a 
cubic GaAs crystal without the inversion center $\beta_{ijk}=\beta$ if 
all $i,j,k$ are different and zero otherwise. In the limit of large $d$ 
the piezoelectric part dominates over the deformation potential but in 
the experiments \cite{BF98,LTHS00} $d\approx 50A$ and both 
exciton-phonon interaction terms are of the same order of magnitude: 
$\Theta\approx 4\pi ed\beta \approx 10eV$.

Amplitude of phonon emission is given to the lowest order of the 
perturbation theory by the following matrix element:
\begin{equation}\label{M} M^s_{if}(\vec{q})=<f\vec{q}s|H_{ph}|i0>
\end{equation} between initial state of Bose gas $|i0>$ with no phonons
and the final state of Bose gas $|f\vec{q}s>$ with just one phonon 
specified by the momentum: $\vec{q}$ and the polarization $s$. We 
assume that the thermalization of the Bose gas due to particle-particle 
scattering is much faster than the slow cooling due to phonon emission. 
Thus, at any given time $t$ the Bose gas is characterized by an 
effective temperature: $T(t)$ . This temperature defines the total gas 
energy: $E=E(T)$. The Fermi Golden Rule gives the probability of 
phonon emission per unit time and one needs to multiply it by the 
phonon energy: $\omega_s(\vec{q})=cq$, to find the total energy losses:
\begin{equation}\label{Wif} {dE\over dt}=-{2\pi\over\hbar}\sum_{f
\vec{q}s}cq|M^s_{if}(\vec{q})|^2\delta(E_i-E_f-cq). \end{equation} 
Eq.(\ref{Wif}) has to be averaged over Gibbs distribution of the 
initial states with the effective temperature $T(t)$. Both the 
initial and the final states of the Bose gas are calculated in the 
interaction representation (see e.g.\cite{AGD}). Particles are 
confined to the 2D layer and the energy losses are proportional to 
the area of this layer.

In the experiments \cite{BF98,LTHS00} only a relatively low density 
of exciton gas has been achieved. Popov has shown \cite{Popov} that 
for a 2D dilute Bose gas there is a 2D 
Berezinski-Kosterlitz-Thouless phase transition point:
\begin{equation}\label{Tc} T_c={2\pi n\hbar^2\over gm\log{L}}
\end{equation} that separates high-T almost ideal Bose gas phase 
from the low-T superfluid phase. Actually, Popov theory is 
controlled by the large logarithm:
\begin{equation}\label{Log}L\approx -\log(nr_0^2)\approx\log{E_0
\over T_c}\end{equation} where $n$ is the particle density, 
$E_0=\hbar^2/r_0^2m$, $g$ is the particle internal degeneracy 
\cite{Popov}. For a Bose particle: $g=2S+1$, where $S$ is the spin 
of particle. It was shown in Ref.\cite{Vin94} that electron and 
hole spins flip rapidly due to the spin-orbit interaction. Thus, 
$g=4$ for a GaAs exciton.

For 2D dilute non-ideal Bose gas one can distinguish three 
temperature regions. At high temperatures: $T\gg T_c\log{L}$, the 
ideal Bose gas is a good approximation. At intermediate 
temperatures: $gT_c/L\ll T\ll T_c/\log{L}$, an overwhelming amount 
of particles constitute a 2D BEqC with the density:
\begin{equation}\label{CondDens} n_s=n\left(1-{T\over T_c} \right), 
\end{equation} whereas a small fraction of thermal particles have a 
bare dispersion and Bose distribution with the chemical potential: 
$\mu\approx gT_c/L$ \cite{Popov}. At low temperatures: $T\ll\mu$, a 
weak particle-particle interaction is crucial and the Bose gas 
becomes a Bose-liquid with quasiparticle excitations having Bogolubov 
sound like dispersion. The transfer of momentum to an impurity 
becomes here inefficient because sound-like quasiparticles ignore 
point-like impurities. Hot particles with a bare dispersion are 
rare due to an exponentially small Boltzmann factor. In the case of 
excitons in GaAs crystal the intermediate-T region hardly exists 
at all.

\begin{figure} \epsfxsize=4.5in  \centerline{\epsffile{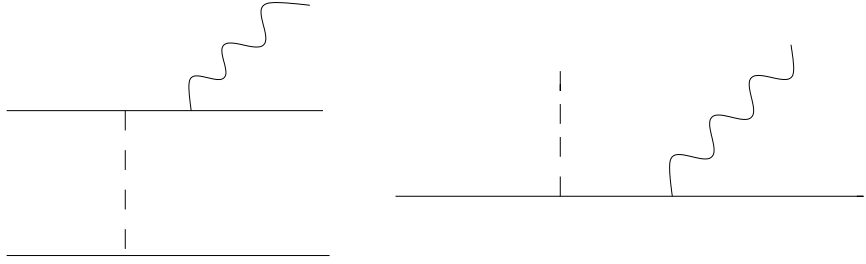}}
\caption{Amplitudes of exciton-exciton scattering (left) and 
impurity scattering (right) accompanied by an emission of phonon - 
shown by a wavy line}\end{figure}

Fig.1 shows two scattering possibilities for a particle accompanied 
by an acoustic phonon emission. The left diagram shows a scattering 
on the second particle and the right diagram shows a scattering on 
impurity. First, we treat this problem in the high-T and 
intermediate-T regions where particles have the bare dispersion: 
$\epsilon=\vec{p}^2/2m$. 2D scattering amplitudes in both cases 
are isotropic and depend only on the total kinetic energy in the 
center of mass frame: ${\cal E}$ (in the impurity case ${\cal E}=
\epsilon$) in the limit ${\cal E}\rightarrow 0$:
\begin{equation}\label{AmplitudeF} F({\cal E})=-{2\pi\hbar^2\over 
m^*}\left(\log{\hbar^2\over{\cal E} r_0^2m}\right)^{-1}, 
\end{equation} where $m^*=m/2$ for particle-particle scattering and 
$m^*=m$ for the impurity scattering (see e.g. \cite{Popov}). Both 
particle-particle interaction line and the impurity line on Fig.1 
correspond to the scattering amplitude $F({\cal E})$. The total 
amplitude of assisted phonon emission is universal in both cases 
and is given by the following matrix element:
\begin{equation}\label{IAPEM} \displaystyle M_{if}^{s}(\vec{q})=C
\left(F({\cal E}){1\over cq}+F({\cal E}-cq){1\over -cq}\right)e^s_i
\Gamma_i(\vec{q})\sqrt{q\over 2\rho c}, \end{equation} where $C=2$ 
for the particle-particle scattering and $C=1$ for the impurity 
scattering. We neglect the phonon momentum $\vec{q}_{||}$ transfer 
to the particle because in the low temperature limit: $q_{||}\ll 
p,p'\ll cm$.

Plugging Eqs.(\ref{IAPEM},\ref{GammaPiez}) into Eq.(\ref{Wif}) and 
taking the integral over the final state of the Bose gas we obtain 
the total energy losses per unit time. In the high-T region we get:
\begin{equation}\label{Prob} {dE\over dt}=-{2\pi c^3\over\hbar 
T^2_{x-ph}}\int\left({1\over\log(E_0/\epsilon)}-{1\over\log(E_0/
\epsilon')}\right)^2 K(\epsilon,T)\delta(\epsilon-\epsilon'-cq)d
\epsilon d\epsilon'{d^3\vec{q}\over(2\pi)^3}, \end{equation} where 
$K(\epsilon,T)=2AnN(\epsilon)(1+1/g)$ in the case of particle-particle 
scattering and $K(\epsilon,T)=An_{imp}N(\epsilon)/2$ in the case of 
impurity scattering. $A$ is the total area of bilayer, $n_{imp}$ is 
the areal density of impurities and
\begin{equation}\label{BE} N(\epsilon)={1\over\exp\left((\epsilon-\mu)
/T\right)-1} \end{equation} is the Bose-Einstein occupation number. 
It is also convenient to define a characteristic exciton-phonon 
temperature in the case of GaAs:
\begin{equation}\label{Texph} T_{x-ph}=\sqrt{\rho c^5\hbar^3\over
\Theta^2+(4\pi de\beta)^2/15}\approx 5^{\circ}K. \end{equation} 
Combining the particle-particle and impurity contributions and using 
the ideal gas equation of state: $E(T)=AnT$, we find the total 
cooling rate:
\begin{equation}\label{Texex} {dT\over dt}=-(4(1+1/g)n+n_{imp})
{85\over 9m}{\hbar\over L^4}{T^3\over T_{x-ph}^2}. \end{equation}
Integrating the Eq.(\ref{Texph}) we get the temperature relaxation 
law: $T(t)\sim 1/\sqrt{t}$.

In the intermediate-T region at $g=1$ the cooling rate is enhanced 
by the stimulated scattering into the BEqC final states:
\begin{equation}\label{TI} {dT\over dt}=-n_s\left(64n_s(1-{\zeta(3)
\over\zeta(2)})+n_{imp}{\zeta(3)\over\zeta(2)}\right){2\pi\hbar^3
\over m^2}{\log^2L\over L^4}{T^2\over T_{x-ph}^2}. \end{equation}

In the low-T region the thermodynamic equation of state reads: $E(T)
=A\zeta(3)T^3/\pi s^2$, where $s$ is the Bogolubov sound velocity:
$s=\sqrt{\mu/m}$. In order to calculate the energy losses we apply a 
Bogolubov unitary transformation to the Hamiltonian (\ref{Ex-PhHam}):
\begin{equation}\label{Ham1} H_{ph}=-\sum_{\vec{p}\vec{q}}{\mu\over 
2\epsilon(\vec{p})}\psi^+(\vec{p})\psi^+(-\vec{p}+\vec{q})\Gamma_i
(\vec{q})u_i(\vec{q})+c.c. \end{equation} This Hamiltonian allows an 
emission of a phonon. The cooling rate in this case is also enhanced 
by the condensate stimulation:
\begin{equation}\label{TI2} {dT\over dt}=-{\mu^2\over\hbar\pi 
T^2_{x-ph}}\left(1-{\zeta(4)\over\zeta(3)}\right) T^2.\end{equation}

\begin{figure} \epsfxsize=4.5in
\centerline{\epsffile{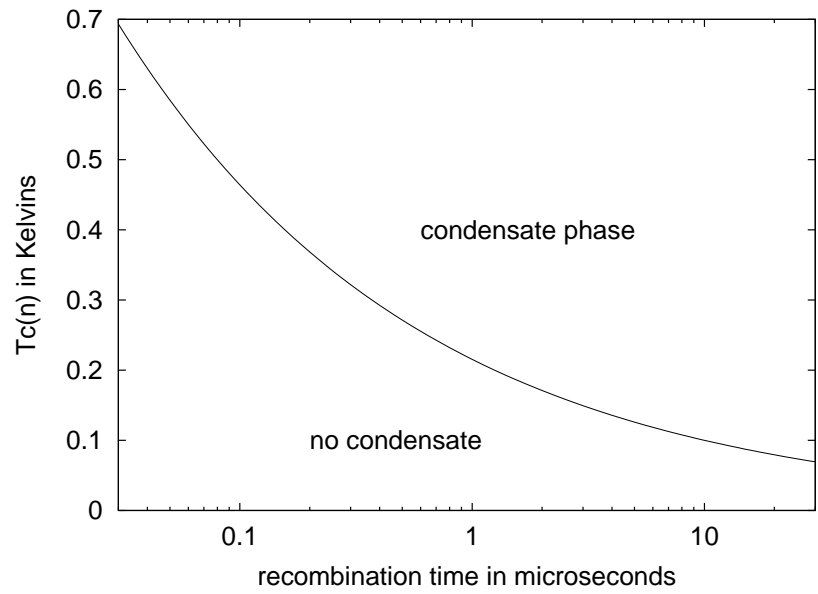}}
\caption{Kinetic phase diagram showing the posibility to reach a Bose 
Einstein quasi-condensate phase in the GaAs indirect exciton bilayer.}
\end{figure}

For experimental realization of an exciton cooling the most relevant 
is Eq.(\ref{Texex}). Integrating it and inserting $L\approx 6$ and 
$n\gg n_{imp}$ we find the overall cooling time $t_c$ required to 
reach the phase transition point. Here one can distinguish two cases: 
i) cooling of exciton gas with constant density e.g. sustained by 
photoexcitation and ii) cooling of decaying exciton gas with 
$n(t)=n_0\exp(-t/\tau_r)$, where $\tau_r$ is the exciton 
recombination time. We find:
\begin{equation}\label{TtLaw} t_c={C\hbar T^2_{x-ph}\over T_c
(n)^3}, \end{equation} where $T_c(n)$ is the BKT temperature as a 
function of the exciton density $n$ (\ref{Tc}), $C\approx 10$ in the 
case i) and $C\approx 30$ in the case ii). Note that $C$ does not 
depend on exciton mass $m$ and in the case ii) the best condition 
for reaching $T_c$ occurs at $t=\tau_r/3$. Eq.(\ref{TtLaw}) defines a 
frontier in the bilayer parameter space: $(n,\tau_r)$ or equivalently 
$(T_c(n),\tau_r)$, separating the two kinetic phases - one that can 
condensate and the second that remains above $T_c$ during the exciton 
life time $\tau_r$. Fig.2 shows this borderline for the case of a 
thin GaAs bilayer.

This work was supported by RFFI and INTAS.


\begin{thebibliography}{22}
\bibitem{BF98} L.V. Butov, A. Imamoglu, A.V. Mintsev, K.L. Campman 
and A.C. Gossard, Phys.Rev.B {\bf 59}, 1625 (1999); L.V. Butov and 
A.I. Filin, Phys.Rev.B {\bf 58}, 1980 (1998)
\bibitem{LTHS00} A.V.Larionov, V.B.Timofeev, I.Hvam and R.Soerensen, 
JETP {\bf 117}, 1255 (2000)
\bibitem{ILH98} A.L.Ivanov, P.B.Littelwood and H.Haug, Phys.Rev.B 
{\bf 59} 5032 (1999)
\bibitem{LR97} Yu.E.Lozovik and A.M.Ruvinskii, JETP {\bf 85}, 979 
(1997)
\bibitem{GL} V.F.Gantmakher and Y.B.Levinson, Scattering of charge 
carriers in metals and semiconductors (Nauka Moscow 1984)
\bibitem{AGD} A.Abrikosov, L.P.Gor'kov and I.E.Dzyaloshinski, Method 
of Quantum Field Theory in Statistical Physics, Nauka Moscow (1962)
\bibitem{Popov} P.N.Brusov and V.N.Popov, Superfluidity and Collective
Properties of Quantum Liquids (Nauka Moscow 1988);
D.S. Fisher and P.C. Hohenberg, Phys.Rev.B {\bf 37}, 4936 (1988)
\bibitem{Vin94} A.Vinattieri, J.Shah, T.C.Damen, D.S.Kim, L.N.Pfeifer,
M.Z.Maialle and L.J.Sham, Phys.Rev.B {\bf 50}, 10868 (1994)
\end{thebibliography}
\end{document}